# The discrete Lamb problem: Elastic lattice waves in a block medium

## N.I. Aleksandrova

*N.A. Chinakal Institute of Mining, Siberian Branch, Russian Academy of Sciences, Krasnyi pr. 54, Novosibirsk, 630091, Russia*
*Novosibirsk State University, ul. Pirogova, Novosibirsk, 630090, Russia*
*e-mail: alex@math.nsc.ru*

## Highlights

• An asymptotic behavior of perturbations for the transient Lamb problem for a point step load in a discrete medium is obtained.
• Numerical examples are given to demonstrate the accuracy of our analytical solutions.
• A new approximate representation of the Lommel function $s_{0,k}$ for $k >> 1$ in terms of the Scorer function Gi is obtained.

## Abstract

We study the propagation of transient waves under the action of a vertical step point load on the surface of a half-space filled by a block medium. The block medium is modeled by a square lattice of masses connected by springs in the directions of the axes *x*, *y*, and in the diagonal directions. The problem is solved by two methods. Analytically, we obtain asymptotic solutions in the vicinity of the Rayleigh wave at large time intervals. Numerically, we obtain a solution for any finite time interval. We compare these solutions with each other and with the solution to the Lamb problem for an elastic medium.

## 1. Introduction

In [1], M.A. Sadovskiy has shown that, in the study of rock masses, it is necessary to take into account their block-hierarchical structure. As noted in [1,2], the structure of a block medium is the cause of various dynamic phenomena that are absent in a homogeneous medium and, therefore, cannot be described by its models. Among those dynamic phenomena we distinguish the propagation of the pendulum waves having a low velocity of spread and long length under a pulse action in rock masses with the block structure. Some peculiarities of the pendulum waves were studied on one-dimensional models of the block media in [3,4]. In these papers we calculated the wave motion in a chain of elastic rods, separated by compressible intermediate layers, and showed that a low-frequency disturbance caused by the pulse action is well described in the model of "rigid blocks and viscoelastic layers". The same approach was used in [5,6] for the description of the dynamic behavior of a two-dimensional block medium in which rigid blocks are assumed to be of a rectangular shape. A simplified model of a block medium can be obtained if we treat the blocks as point masses connected by springs. In this case, a block medium can be represented as a lattice of masses connected with each other by springs. In [7 – 10], different versions of this approach were used in the plane and anti-plane settings.

For the first time, the problem of the dynamic impact of a vertical point force applied to the boundary of an elastic half-space was considered by H. Lamb [11]. In that work he obtained integral representations for the displacements on the boundary of the half-space and showed that, on the surface of the half-space, the Rayleigh waves [12] propagate along with the P- and S-waves. A similar integral representation of a solution to the Lamb problem, both two-dimensional and three-dimensional, was obtained in [13] by the method of functional-invariant solutions. In [14], L. Cagniard obtained a solution to the three-dimensional Lamb problem using the method of integral transformations. To calculate the inverse integral transforms, he proposed a method that now bears his name. Later that method was developed and generalized in [15]. Without going into details, we mention that the Lamb problem and Rayleigh waves were considered in many books, see, e.g., [16–21], and papers, see, e.g., [22,23], devoted to the issues of wave propagation. In most of these books and papers the Cagniard – De Hoop method is used.

The static setting of the plane Lamb problem, when the load does not depend on time, is called the Flamant problem. In this setting the displacement field is given in [24]. A solution for a similar problem for the stress field is given in [18]. In [18] an approximate solution is also obtained for the stationary Lamb problem for displacements on the surface of the half-space under the action of a harmonic point load. The stationary Lamb problem is also considered in [25] within the framework of the Cosserat continuum.

In the transient setting, analytical solutions to the plane Lamb problem for elastic media were considered in [16,17,19–23]. In [26,27], asymptotic solutions are obtained in the vicinity of the quasi-front for the case of the action of an instantaneous point force on a semi-infinite plate; the method of contour integrals is used and the solution is represented in the integral form.

In the transient setting, numerical solutions to the Lamb problem for elastic media were considered in [28], using the finite-element method for plane and spatial problems for the case of harmonic loading and in [29], using the finite-difference method for the plane problem for the case of distributed loading on the free surface.

For a discrete medium composed of a lattice of point-masses that simulates a continuous elastic medium, the Lamb problem about the propagation of disturbances was solved in [7], where a solution was obtained for 2D- and 3D-lattices in the form of multiple integrals, which are similar to the integral representations of the Bessel functions. Thus, in [7,29], various discretizations of the elastic medium for the Lamb problem were considered. Unlike [7,29], in this paper we obtain both a finite-difference solution and an analytical solution to the Lamb problem for a discrete medium and compare these solutions with each other and with the solution for an elastic medium given in [16,23].

## 2. Setting of the problem

In the present paper, we study the transient plane Lamb problem on the impact of a vertical point load on the boundary of a half-plane filled by a block medium as it is shown in Fig. 1, where *u* is the horizontal displacement, *v* is the vertical displacement, *n, m* are the indices of the masses in the directions *x, y,* and $P_0$ is the amplitude of the point load. The block medium is modeled by a uniform two-dimensional lattice consisting of the masses connected by springs in the directions of the axes *x, y*, and in the diagonal directions (Fig. 2). This is a special case of the model proposed in [8].

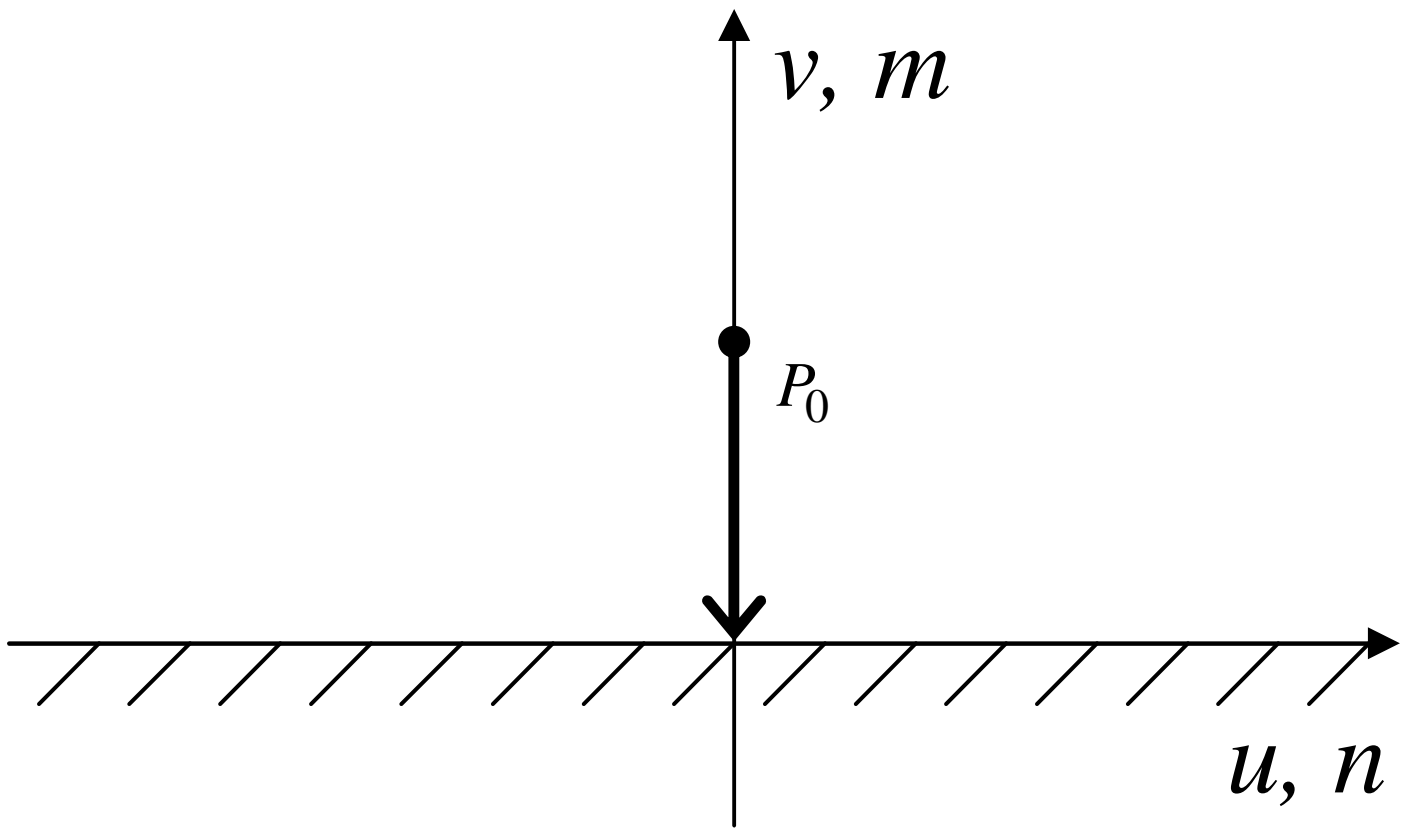

**Fig. 1.** Setting of the problem.

The equations of the motion of the mass with indices *n, m*, located apart from the boundary, are as follows:

$$\begin{aligned} M\ddot{u}_{n,m} &= k_1(u_{n+1,m} - 2u_{n,m} + u_{n-1,m}) + k_2(u_{n+1,m+1} + u_{n-1,m-1} + u_{n+1,m-1} \\ &\quad + u_{n-1,m+1} - 4u_{n,m})/2 + k_2(v_{n+1,m+1} + v_{n-1,m-1} - v_{n-1,m+1} - v_{n+1,m-1})/2, \\ M\ddot{v}_{n,m} &= k_1(v_{n,m+1} - 2v_{n,m} + v_{n,m-1}) + k_2(u_{n+1,m+1} + u_{n-1,m-1} - u_{n+1,m-1} \\ &\quad - u_{n-1,m+1})/2 + k_2(v_{n+1,m+1} + v_{n-1,m-1} + v_{n-1,m+1} + v_{n+1,m-1} - 4v_{n,m})/2. \end{aligned} \tag{1}$$

Here *M* is the mass of a block; $k_1$ is the spring stiffness in the directions of the axes *x, y;* $k_2$ is the spring stiffness in the diagonal directions. The initial conditions are supposed to be zero:

$$u_{n,m} = \dot{u}_{n,m} = v_{n,m} = \dot{v}_{n,m} = 0.$$

We assume that the numbers $m < 0$ correspond to the half-plane. Free boundary corresponds to $m = 0$, see Fig. 1. The equations of the motion of the blocks on the boundary are as follows:

$$\begin{aligned} \gamma M\ddot{u}_{n,0} &= k_1(u_{n+1,0} - 2u_{n,0} + u_{n-1,0}) + k_2(u_{n-1,-1} + u_{n+1,-1} - 2u_{n,0})/2 + k_2(v_{n-1,-1} - v_{n+1,-1})/2, \\ \gamma M\ddot{v}_{n,0} &= k_1(v_{n,-1} - v_{n,0}) + k_2(u_{n-1,-1} - u_{n+1,-1})/2 + k_2(v_{n-1,-1} - 2v_{n,0} + v_{n+1,-1})/2 + P_0\delta_{0n}H(t), \end{aligned} \tag{2}$$

where $H(t)$ is the Heaviside step-function, $\delta_{0n}$ is the Kronecker delta, $P_0$ is the amplitude of the load, and $\gamma$ is the ratio of the mass of a block on the boundary to the mass of a block inside the half-plane ($\gamma > 0$).

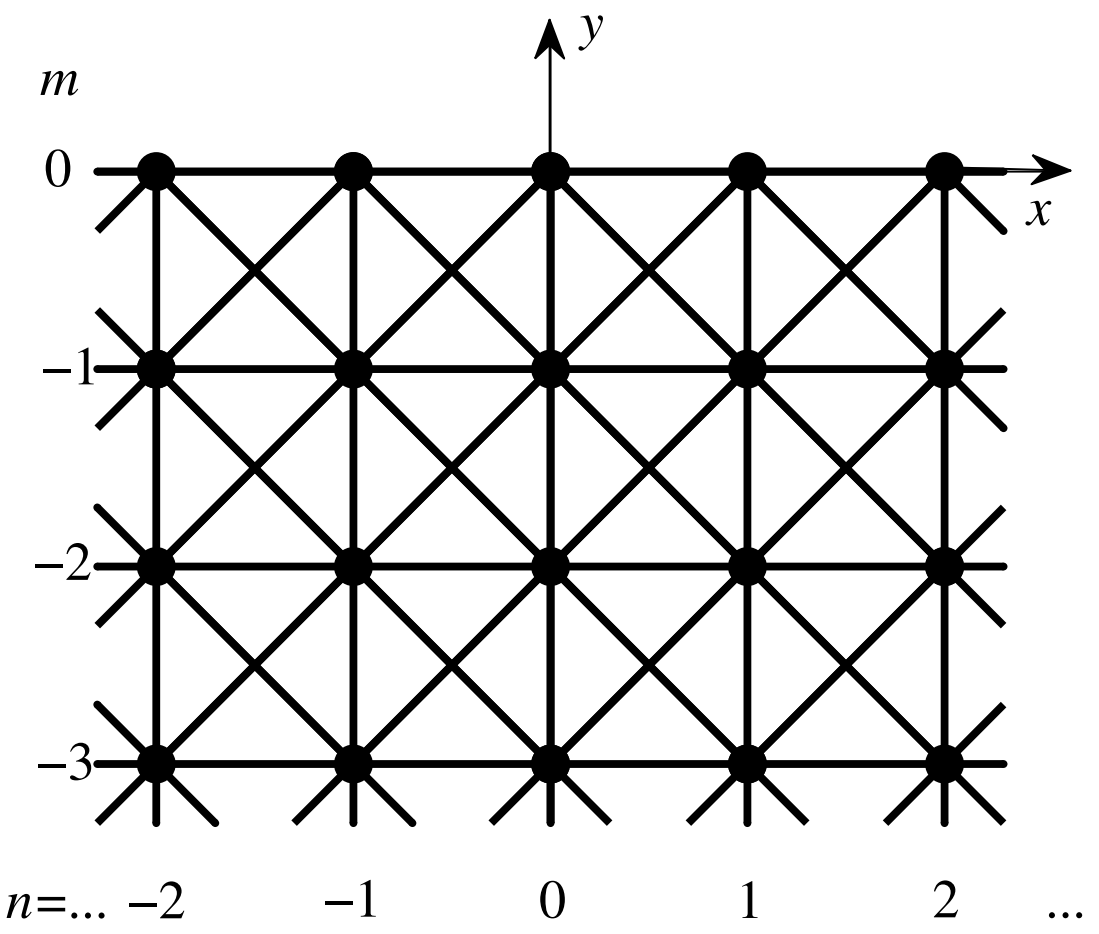

**Fig. 2.** Scheme of connections of the masses by springs in the square lattice.

Let $l$ be the length of a spring in the direction of the axis *x* or *y*. In Eqs. (1), replace the differences in parenthesis by their differential approximations (e.g., $u_{n+1,m} - 2u_{n,m} + u_{n-1,m} \approx$ $\approx l^2 \frac{\partial^2 u}{\partial x^2} + \frac{l^4}{12} \frac{\partial^4 u}{\partial x^4} + \ldots$) and, assuming $l \to 0$, omit the summands of the orders higher than $l^2$. This transformation corresponds to the passage from the block medium to the homogeneous elastic medium. As a result we see that, for $l$ small enough, Eqs. (1) turn into the equations of the orthotropic elasticity theory:

$$M\ddot{u}_{n,m} = l^2 \left[ (k_1 + k_2) \frac{\partial^2 u}{\partial x^2} + k_2 \frac{\partial^2 u}{\partial y^2} + 2k_2 \frac{\partial^2 v}{\partial x \partial y} \right], \tag{3}$$

$$M\ddot{v}_{n,m} = l^2 \left[ (k_1 + k_2) \frac{\partial^2 v}{\partial y^2} + k_2 \frac{\partial^2 v}{\partial x^2} + 2k_2 \frac{\partial^2 u}{\partial x \partial y} \right].$$

If $k_1 = 2k_2$, Eqs. (3) are the equations of a continuous medium and describe the plane stress state of an isotropic elastic medium with the Poisson ratio $\sigma = 1/3$.

Below we always assume $k_1 = 2k_2$. In this case the velocities $c_p$ and $c_s$ of longitudinal and shear long waves in the block medium are independent of the direction of wave propagation (i.e. the lattice is ''isotropic'') and can be found from the formulas [10]:

$$c_p = l\sqrt{3k_1/(2M)}, \qquad c_s = l\sqrt{k_1/(2M)}. \tag{4}$$

Note that, from any single node, the perturbations propagate only in the directions *x, y* and the diagonal directions. But if we consider the lattice as a whole, the disturbances propagate in all directions from the point of impact (i.e., there are no preferred directions of wave propagation; see Fig. 14).

The mass of the blocks and the length of the springs are taken as the units: $M = 1$, $l = 1$. We assume $k_1 = 3/4$. The value of $k_1$ is chosen so that the velocity of longitudinal infinitely long waves in the lattice is equal to the velocity of longitudinal waves in an isotropic elastic medium in the case of plane stress with the Poisson ratio $\sigma = 1/3$: $c_p = l\sqrt{3k_1/(2M)} = \sqrt{E/(\rho(1-\sigma^2))}$, where $E$ is the Young modulus and $\rho$ is the density of the elastic medium. The velocity $\sqrt{E/\rho}$ is taken as unity.

The Rayleigh equation for the elastic medium is as follows [12]:

$$\left(2-\frac{c^2}{c_s^2}\right)^2 - 4\sqrt{\left(1-\frac{c^2}{c_p^2}\right)\left(1-\frac{c^2}{c_s^2}\right)} = 0,$$

where $c_s$ is the velocity of shear wave in the elastic medium. In the case of plane stress, we have $c_s^2/c_p^2 = (1-\sigma)/2$. Assuming as before $\sigma = 1/3$, we find that the Rayleigh equation gives the following value for the velocity of the Rayleigh wave in the elastic medium $(0 < c_R < c_s)$:

$$c_R = \sqrt{\frac{E\left(3-\sqrt{3}\right)}{4\rho}} = \frac{\sqrt{3-\sqrt{3}}}{2} \approx 0.563\ldots. \qquad (5)$$

### 3. Analytical solution

In order to construct an analytical solution to the Lamb problem for the block medium, we apply the Laplace transform (of parameter *p*) with respect to time *t* (denoted by the superscript *L*), and the discrete Fourier transform (of parameter *q*) with respect to *n* (denoted by the superscript $F$):

$$f^L(p) = \int_0^\infty f(t)e^{-pt}dt, \qquad g^F(q) = \sum_{n=-\infty}^{n=\infty} g_n e^{iqnl}.$$

Using the standard approach for obtaining the solution to the Lamb problem for an elastic medium and omitting intermediate computations, we obtain the Laplace–Fourier transform of the solution to the Lamb problem for the block medium:

$$u_m^{LF} = CA_1e^{\alpha_1 m} + DA_2e^{\alpha_2 m}, \qquad v_m^{LF} = CB_1e^{\alpha_1 m} + DB_2e^{\alpha_2 m}, \qquad (6)$$

where

$$C = -\frac{2P_0a_{12}}{pk_1\Delta}, \quad D = \frac{2P_0a_{11}}{pk_1\Delta}, \quad \text{and} \quad \Delta = a_{11}a_{22} - a_{12}a_{21},$$

$$a_{11} = \cos(ql/2)\left[\tilde{M}\left(H(2\gamma-1)+F\right)+4A(H+F)\right], \quad a_{12} = \sin(ql/2)\left[G\left(\tilde{M}(2\gamma-1)+4\right)+(\tilde{M}+4A)J\right],$$

$$a_{21} = i\sin(ql/2)\left[F\tilde{M}(2\gamma-1)+\tilde{M}H+4H\right], \quad a_{22} = i\cos(ql/2)\left[G\tilde{M}(2\gamma-1)+J\left(\tilde{M}+4A\right)\right],$$

$$A_1 = H\cos(ql/2),\quad B_1 = iF\sin(ql/2),\quad A_2 = J\sin(ql/2),\quad B_2 = iG\cos(ql/2),$$

$$e^{-\alpha_1} = \tilde{M} + 1 + 2A - FH\,,\quad e^{-\alpha_2} = \left(\tilde{M} + 3 + 2A - GJ\right)/(3-4A),\quad \tilde{M} = Mp^2/k_1\,,$$

$$H = \sqrt{\tilde{M} + 2A},\quad F = \sqrt{\tilde{M} + 2 + 2A},\quad J = \sqrt{\tilde{M} + 6 - 2A},\quad G = \sqrt{\tilde{M} + 6A},\quad A = \sin^2(ql/2).$$

The dispersion equation $\Delta = 0$ is the Rayleigh equation for the block medium. Assume $p = iqc_R$ and $q \to 0$. After omitting the summands of the orders higher than $q^4$ in the dispersion equation $\Delta = 0$, we obtain the following equation for the velocity of the Rayleigh waves in the block medium $c_R$:

$$2iq^4 l^4\left[\sqrt{3}(1-\tilde{c}^2)^2 - \sqrt{(1-2\tilde{c}^2)(3-2\tilde{c}^2)}\right] = 0\,,$$

where $\tilde{c}^2 = c_R^2 M/(k_1 l^2)$. The solution to this equation has the same value (5) as that for the elastic medium in the case of the plane stress state ($\sigma = 1/3$) and is independent of the coefficient $\gamma$ which determines the weight of blocks on the boundary of the block medium:

$$c_R = l\sqrt{\frac{k_1}{M}\left(1 - \frac{1}{\sqrt{3}}\right)} = \frac{\sqrt{3-\sqrt{3}}}{2} \approx 0.563... \tag{7}$$

Put $p = i\omega$ and $ql = \pi$ in the dispersion equation $\Delta = 0$. As a result we obtain the following equation for the resonance frequency $\omega$ of the short-wave perturbations in the Rayleigh wave:

$$(2\gamma - 1)\,M\omega^2/k_1 - \sqrt{\left(M\omega^2/k_1 - 4\right)\left(M\omega^2/k_1 - 2\right)} = 0\,.$$

The smallest positive root of this equation is given by the formula:

$$\omega_1 = \sqrt{\frac{k_1}{M}}\begin{cases}\sqrt{4/3}, & \gamma = 1,\\[4pt] \dfrac{3 - \sqrt{9 - 32\gamma(1-\gamma)}}{4\gamma(1-\gamma)}, & \gamma > 0,\quad \gamma \neq 1.\end{cases} \tag{8}$$

Taking into account that $k_1 = 3/4, M = 1$, we find from (8) that (i) for $\gamma = 1/2$, the frequency $\omega_1$ is maximal and is equal to $\sqrt{3/2} \approx 1.225...$ and (ii) for $\gamma = 1$, the frequency $\omega_1$ is equal to 1.

In Fig. 3 we show the plots of the phase and group velocities and the wave frequency versus the wave number $q$ corresponding to the first oscillation mode that are obtained numerically from the equation $\Delta = 0$ for the different values of the parameter $\gamma$. The analysis of the lines in Figs. 3(a) and 3(b) shows that, for $\gamma = 1$, the infinitely long waves propagate without dispersion, i.e., the phase and group velocities are equal to each other, while, for the other values of $\gamma$, there is a strong dispersion on the boundary of the block medium.

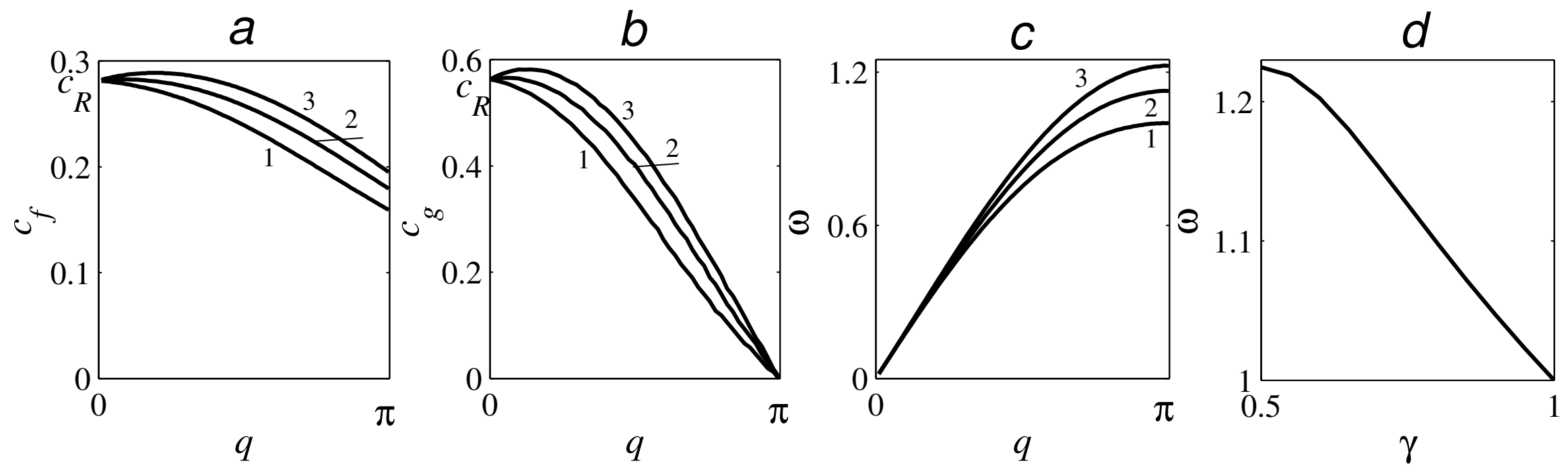

**Fig. 3.** (a, b, c) The dependence of the phase velocity, group velocity, and frequency versus the wave number $q$: lines 1 – $\gamma = 1$, lines 2 – $\gamma = 3/4$, lines 3 – $\gamma = 1/2$. (d) The dependence of the short wave frequency versus the parameter $\gamma$.

From formulas (6) we obtain the following Laplace–Fourier transform of the solution on the boundary ($m = 0$) of the half-plane filled with the block medium in the case $\gamma = 1$:

$$u_0^{LF} = \frac{P_0 \sin(ql)[-(4+\tilde{M})GH + (\tilde{M}+4A)FJ]}{p\Delta k_1},$$
$$v_0^{LF} = \frac{2iP_0[(\tilde{M}+4A)\{-JAF + (H+F)G(1-A)\} - (4+\tilde{M})GAF]}{p\Delta k_1}. \tag{9}$$

Formally, the solution to the transient problem can be written as follows:

$$u_{n,0}(t) = \frac{l}{4\pi^2 i} \int_{-\pi/l}^{\pi/l} \int_{i\alpha-\infty}^{i\alpha+\infty} u_0^{LF}(p,q) e^{pt - iqnl} dp dq.$$

A similar formula holds true for the vertical displacement $v$.

Note that it is not possible to find explicitly the inverse transforms of solution (9). We will seek an asymptotic solution for the long-wave perturbations ($|q| << 1$) propagating with the velocity of the Rayleigh waves ($p \sim iqc_R$) at large time intervals ($t \to \infty \Rightarrow p \to 0$). The asymptotic behavior of the denominator in (9) is as follows:

$$p\Delta k_1 \approx ipk_1 \left[\sqrt{12}(\tilde{M}+4A)^2 - 16A\sqrt{\tilde{M}+2A}\sqrt{\tilde{M}+6A}\right].$$

Introduce the notation:

$$\Delta_1 \approx i\left[\sqrt{12}(\tilde{M}+4A)^2 + 16A\sqrt{\tilde{M}+2A}\sqrt{\tilde{M}+6A}\right].$$

Multiply the numerator and denominator of each of the functions in (9) by $\Delta_1$. It can be shown that, for $p \to 0$ and $|q| << 1$, the following asymptotic expansion holds true:

$$\Delta\Delta_1 \approx -\frac{256\tilde{M}A^2 l^2}{c_R^2\sqrt{3}}\left(p^2 + 4\frac{c_R^2}{l^2}A\right). \tag{10}$$

Expand the numerators of the expressions (9) in powers of $p$ and $\sin(ql/2)$ and omit the terms of degree higher than 2. Combining the resulting asymptotic formula with formula (10), we obtain the following asymptotic behaviors of the Laplace–Fourier transforms of the velocities of the displacements in the vicinity of the Rayleigh wave:

$$\dot{u}_0^{LF} \approx \frac{ic_R^2 P_0 \sin(ql)}{2l^2 k_1 \left(p^2 + 4c_R^2 A/l^2\right)}, \quad \dot{v}_0^{LF} \approx \frac{P_0 \lambda \sin(ql/2)}{M\left(p^2 + 4c_R^2 A/l^2\right)}, \quad \lambda = \left(1 - \frac{1}{\sqrt{3}}\right)\sqrt{\frac{2}{\sqrt{3}} + 1} \approx 0.62\ldots.$$

Inverting the Laplace transform, we obtain the following solutions:

$$\dot{u}_0^F(t) \approx \frac{iP_0 c_R}{2lk_1}\cos\left(\frac{ql}{2}\right)\sin\left[2\frac{c_R t}{l}\sin\left(\frac{ql}{2}\right)\right], \quad \dot{v}_0^F(t) \approx \frac{P_0 l\lambda}{2Mc_R}\sin\left[2\frac{c_R t}{l}\sin\left(\frac{ql}{2}\right)\right].$$

Inverting the discrete Fourier transform with the help of [30], we obtain the asymptotic behaviors of the velocities of displacements on the boundary of the half-plane filled with the block medium in the vicinity of the Rayleigh wave:

$$\dot{u}_{n,0}(t) \approx \frac{P_0 c_R^2}{\pi l k_1}\int_0^{\pi/2} \cos(z)\sin(2zn)\sin\left(2\frac{c_R t}{l}\sin z\right)dz \approx \frac{P_0 n}{2k_1 t} J_{2n}\left(2\frac{c_R t}{l}\right), \tag{11}$$

$$\dot{v}_{n,0}(t) \approx \frac{P_0 l\lambda}{\pi M c_R}\int_0^{\pi/2} \cos(2zn)\sin\left(2\frac{c_R t}{l}\sin z\right)dz \approx \frac{P_0 l\lambda}{\pi M c_R} s_{0,2n}\left(2\frac{c_R t}{l}\right). \tag{12}$$

Here $J_{2n}$ is the Bessel function of the first kind and $s_{0,2n}$ is the Lommel function [30].

According to [30, 31], for $k >> 1$ the following approximation is valid:

$$J_k(\beta t) \approx \frac{1}{(\beta t/2)^{1/3}}\mathrm{Ai}\left(\frac{k - \beta t}{(\beta t/2)^{1/3}}\right), \quad \mathrm{Ai}(\kappa) = \frac{1}{\pi}\int_0^\infty \cos\left(\kappa z + \frac{z^3}{3}\right)dz, \tag{13}$$

where $\mathrm{Ai}(\kappa)$ is the Airy function. Substituting (13) into (11) yields:

$$\dot{u}_{n,0} \approx \frac{P_0 nl}{4k_1 t}\frac{\mathrm{Ai}(\kappa)}{(\alpha t)^{1/3}}, \qquad \kappa = \frac{nl - c_R t}{(\alpha t)^{1/3}}, \qquad \alpha = \frac{l^2 c_R}{8}. \tag{14}$$

In formula (14) and below, we assume that $n \geq 0$. If $n < 0$, the solution may be found from the following symmetry conditions: $u_{-n,m} = -u_{n,m}$, $v_{-n,m} = v_{n,m}$.

It follows from (14) that the amplitude of the velocity of the horizontal displacement in the vicinity of the front of the Rayleigh wave $nl = c_R t$ decreases as $t^{-1/3}$, when $t \to \infty$ or as $n^{-1/3}$, when $n \to \infty$. Under the same conditions the width of the quasi-front, characterized by the parameter $\kappa$, grows as $t^{1/3}$ or $n^{1/3}$.

Using (11), we find the asymptotic behavior of the horizontal displacement:

$$u_{n,0} \approx \frac{P_0}{2\pi k_1}\int_0^{\pi/2} \frac{\cos(z)\sin(2zn)}{\sin z}\left[1 - \cos\left(2\frac{c_R t}{l}\sin z\right)\right]dz \approx$$

$$\approx \frac{P_0}{2\pi k_1}\int_0^{\infty}\frac{\sin(2zn)}{z}\left\{1-\cos\left[2\frac{c_R t}{l}z\left(1-\frac{z^2}{6}\right)\right]\right\}dz \approx \frac{P_0}{4k_1}\left[\frac{1}{3}-\int_0^{\kappa}\mathrm{Ai}(y)dy\right]. \quad (15)$$

Using the above described method we cannot derive a formula for the velocity of the vertical displacement $\dot{v}$ that will be similar to (14). However, using the Slepyan method [16] of combined asymptotic ($t \to \infty$) inversion of the integral Laplace–Fourier transforms of long-wave disturbances in the vicinity of the ray $x = c_* t$, where $c_*$ is the velocity of perturbations propagating without dispersion, we can find the asymptotic behavior of the solution $\dot{v}$. Let us present the main ideas of the Slepyan method as applied to $\dot{v}_{n,0}$ in the vicinity of the ray $nl = c_R t$.

In order to obtain the inverse Laplace–Fourier transform on the ray $nl = c_R t$ we make the substitution $p = s + iq(c_R + c')$, $nl = (c_R + c')t$, where $c' \to 0$ and defines the vicinity of the ray $nl = c_R t$, in the integral

$$\dot{v}_{n,0}(t) = \frac{l}{4\pi^2 i}\int_{-\pi/l}^{\pi/l}\int_{i\alpha-\infty}^{i\alpha+\infty}\dot{v}_0^{LF}(p,q)e^{pt-iqnl}dpdq, \quad \text{where} \quad \dot{v}_0^{LF}(p,q) = \frac{P_0\lambda\sin(ql/2)}{M\left(p^2+4c_R^2\sin^2(ql/2)/l^2\right)}.$$

This yields

$$\dot{v}_{n,0}(t) = \frac{l}{4\pi^2 i}\int_{-\pi/l}^{\pi/l}\int_{\alpha-i\infty}^{\alpha+i\infty}\dot{v}_0^{LF}(s+iq(c_R+c'),q)e^{st}dsdq.$$

We expand the numerator and denominator of the function $\dot{v}_0^{LF}(s+iq(c_R+c'),q)$ in the Taylor series in a small neighborhood of the point $q = 0$ as $s \to 0$ and $c' \to 0$:

$$\dot{v}_{n,0} \approx \frac{l}{4\pi^2 i}\int_{-\varepsilon}^{\varepsilon}\int_{\alpha-i\infty}^{\alpha+i\infty}\frac{P_0|q|l\lambda}{4Miqc_R\left(s+iqc'+iq^3\alpha/3\right)}e^{st}dsdq,$$

where $\varepsilon > 0$ is small enough. Successively integrating and taking into account that $c' = (nl - c_R t)/t$, we get

$$\dot{v}_{n,0} \approx -\frac{P_0 l^2\lambda}{4\pi M c_R}\int_0^{\varepsilon}\sin\left(qc't+\frac{q^3\alpha t}{3}\right)dq \approx -\frac{P_0 l^2\lambda}{4Mc_R}\frac{\mathrm{Gi}(\kappa)}{(\alpha t)^{1/3}}, \quad \mathrm{Gi}(\kappa) = \frac{1}{\pi}\int_0^{\infty}\sin\left(\kappa z+\frac{z^3}{3}\right)dz, \quad (16)$$

where $\alpha, \kappa$ are defined in (14) and Gi is a Scorer function [32]. Plots of the functions $\mathrm{Ai}(\kappa), \mathrm{Gi}(\kappa)$ are shown in Fig. 4(a).

Using the same method of Slepyan we get the following asymptotic formulas for $u_{n,0}, \dot{u}_{n,0}$:

$$u_{n,0} \approx \frac{P_0}{4k_1}\left[\frac{1}{3}-\int_0^{\kappa}\mathrm{Ai}(y)dy\right], \quad (17)$$

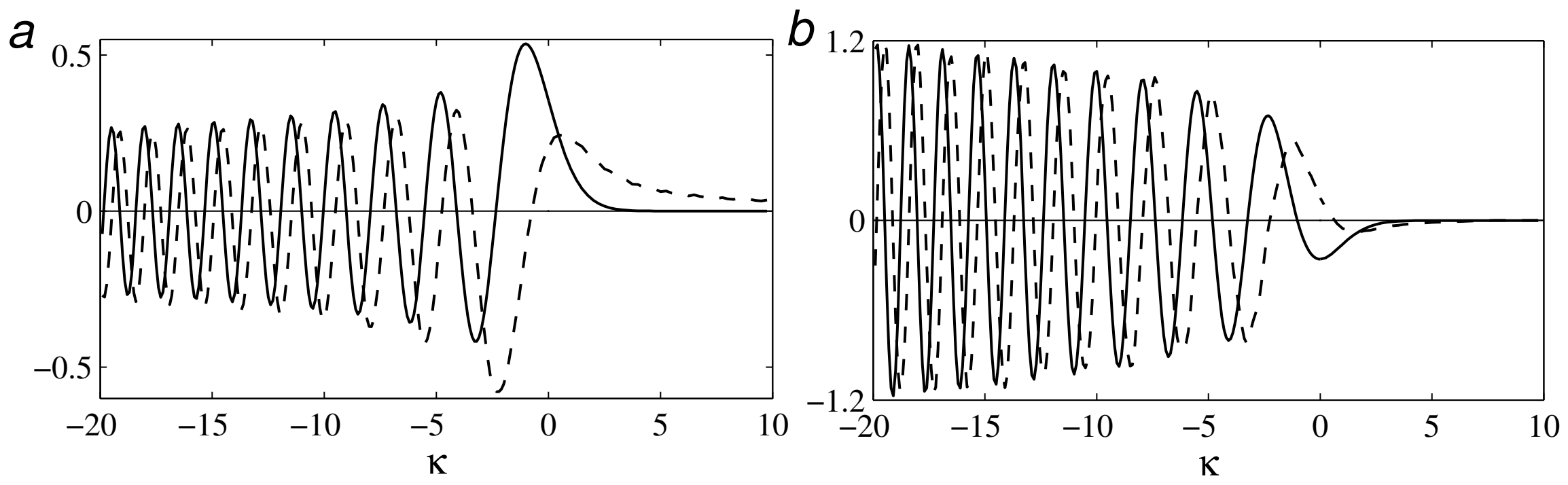

**Fig. 4.** Plots of the functions $\mathrm{Ai}(\kappa), \mathrm{Gi}(\kappa)$ and their derivatives: (a) $\mathrm{Ai}(\kappa)$ – solid line, $\mathrm{Gi}(\kappa)$ – dashed line, (b) $\mathrm{Ai}'(\kappa)$ – solid line, $\mathrm{Gi}'(\kappa)$ – dashed line.

$$\dot{u}_{n,0} \approx \frac{P_0 c_R}{4k_1} \frac{\mathrm{Ai}(\kappa)}{(\alpha t)^{1/3}}. \tag{18}$$

Comparing (15) and (17), we see that the asymptotic behaviors of the horizontal displacement $u$ obtained by the two different methods, are the same. Note that the asymptotic behavior of the velocity of the horizontal displacement $\dot{u}$ (18), obtained by the Slepyan method [16], coincides with (14) only in the vicinity of the quasi-front $nl = c_R t$, while behind the Rayleigh wave, solution (18) decreases slower than (14).

It follows from (16) that the amplitude of $\dot{v}$ in the vicinity of the front of the Rayleigh wave $nl = c_R t$ decreases as $t^{-1/3}$, when $t \to \infty$ or as $n^{-1/3}$, when $n \to \infty$. The width of the quasi-front of the Rayleigh wave, characterized by the parameter $\kappa$, grows as $t^{1/3}$ or $n^{1/3}$. Thus, from (14) and (16), we see that, in the vicinity of the front of the Rayleigh wave, the degrees of decreasing of the amplitudes of $\dot{u}$ and $\dot{v}$ are the same for $t \to \infty$. Similarly, the quasi-front zones of $\dot{u}$ and $\dot{v}$ expand equally.

Comparing (12) and (16), we get the following approximate representation of the Lommel function $s_{0,k}$ for $k >> 1$ in terms of the Scorer function Gi that is similar to (13):

$$s_{0,k}(\beta t) \approx -\frac{\pi}{2(\beta t/2)^{1/3}} \mathrm{Gi}\left( \frac{k - \beta t}{(\beta t/2)^{1/3}} \right), \qquad k = 2n. \tag{19}$$

Differentiating (11), (12), (16), and (18) with respect to time, we obtain the following two asymptotic formulas for each of the acceleration of the displacements for $t \to \infty$:

$$\ddot{u}_{n,0}(t) \approx \frac{P_0 n c_R}{k_1 t l} J'_{2n}\left( 2\frac{c_R t}{l} \right), \tag{20}$$

$$\ddot{v}_{n,0}(t) \approx \frac{2P_0 \lambda}{\pi M} s'_{0,2n}\left( 2\frac{c_R t}{l} \right), \tag{21}$$

$$\ddot{u}_{n,0}(t) \approx -\frac{P_0}{k_1}\left(\frac{c_R t}{l}\right)^{1/3}\frac{n\mathrm{Ai}'(\kappa)}{t^2}, \tag{22}$$

$$\ddot{v}_{n,0}(t) \approx \frac{P_0 l^2 \lambda}{4M}\frac{\mathrm{Gi}'(\kappa)}{(\alpha t)^{2/3}}. \tag{23}$$

Here the prime denotes differentiation with respect to the argument. Plots of the functions $\mathrm{Ai}'(\kappa), \mathrm{Gi}'(\kappa)$ are shown in Fig. 4(b). From (22), (23), we see that the acceleration of the horizontal $\ddot{u}$ and vertical $\ddot{v}$ displacements in the vicinity of the front of the Rayleigh wave decreases as $t^{-2/3}$ (or $n^{-2/3}$), when $t \to \infty$ (or $n \to \infty$). Behind the front of the Rayleigh wave $\ddot{u}$ decreases faster ($\approx t^{-5/6}$) than $\ddot{v}$ ($\approx t^{-2/3}$).

## 4. Comparison of numerical and analytical solutions

In order to determine the limits of applicability of the above analytical solutions, we solve Eqs. (1) and (2) by a finite difference method using an explicit scheme. For the second derivatives with respect to time we use the central difference approximation of the second order of accuracy:

$$\ddot{u}_{n,m} \approx \left(u_{n,m}^{s+1} - 2u_{n,m}^{s} + u_{n,m}^{s-1}\right)/\tau^2, \quad s = 0,1,2,\ldots.$$

Here $\tau$ is the time step of the difference mesh, $u_{n,m}^{s}$ is the value of the displacement $u_{n,m}(t)$ at time $t = s\tau$, $s$ is the number of the time step in the finite difference scheme. The stability condition of the difference scheme depends on the coefficient $\gamma$:

$$\tau \le l\sqrt{\frac{M}{k_1}}\begin{cases}\sqrt{8\gamma/9}, & 0 < \gamma \le 3/4,\\ \sqrt{2/3}, & 3/4 \le \gamma \le 1.\end{cases}$$

Numerical calculations confirm the validity of this stability condition.

Below we present the results of the calculations of disturbances in the Lamb problem for a block medium under the action of the vertical step load at the point with coordinates $n = 0, m = 0$. Since, for $\gamma = 1$, the oscillation frequency of the short-wave perturbations is equal to $\omega_1 = 1$, the length of the half-wave (in time) is equal to $t_0 = \pi/\omega_1 = \pi$. On a single half-wave there should be at least 8-10 time steps in order that the numerical solution describes the short-wave perturbations with an acceptable accuracy. For this reason, we put $\tau = t_0/10 = 0.314$. All of the following numerical calculations were carried out for $\tau = 0.314$, $P_0 = 1$.

In Figs. 5 and 7, the distributions of the displacements and velocities of the displacements along the axis *n* are presented at the moment of time *t*. On those figures, the vertical lines correspond to the coordinates of the quasi-fronts of the longitudinal, shear, and Rayleigh waves (4), (7):

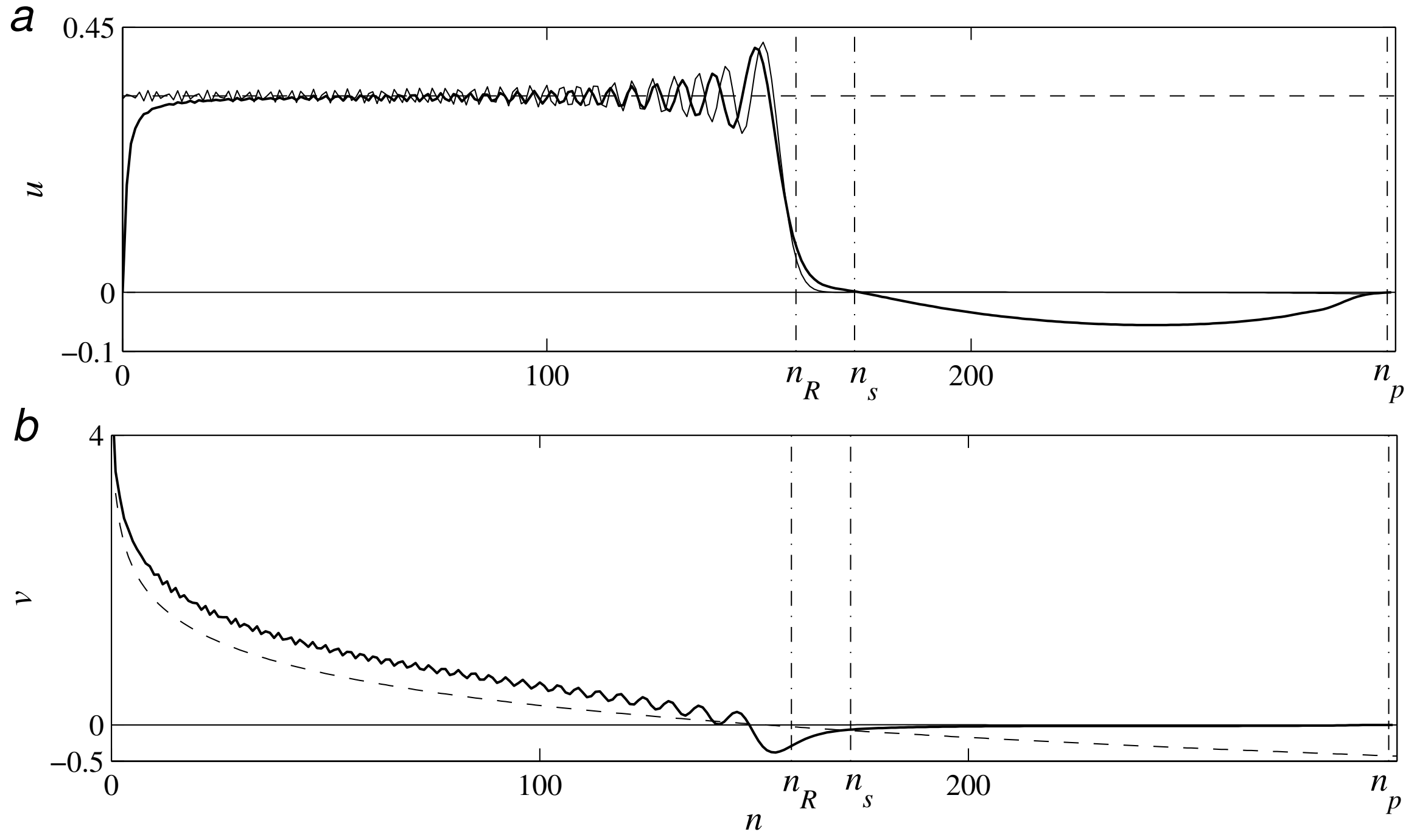

**Fig. 5.** Dependence of displacements on the boundary of the block medium versus coordinate $n$ ($t = 280$, $\gamma = 1$). (a) Horizontal displacement *u,* (b) vertical displacement *v*. Dashed lines – static solution (24). Thick solid lines – the results of numerical calculations. Thin solid line – asymptotic solution (15).

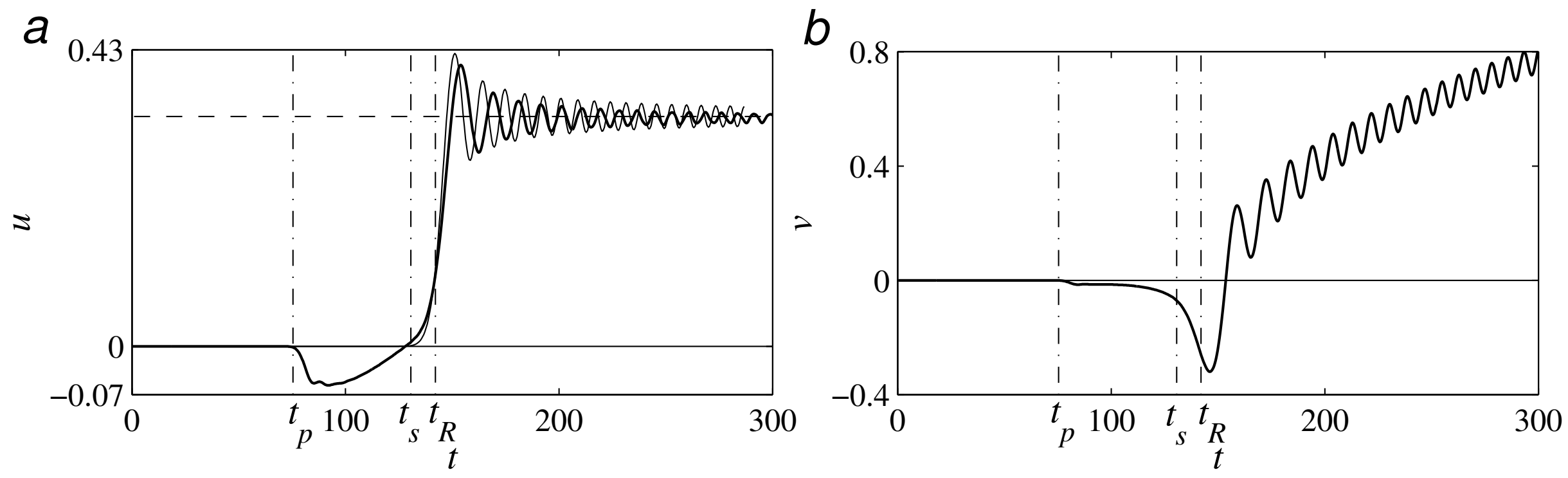

**Fig. 6.** Oscillograms of the displacements at the point $n = 80, m = 0$ ($\gamma = 1$). (a) Horizontal displacement *u,* (b) vertical displacement *v*. Thick solid lines – the results of numerical calculations. Dashed line – static solution (24). Thin solid line – asymptotic solution (15).

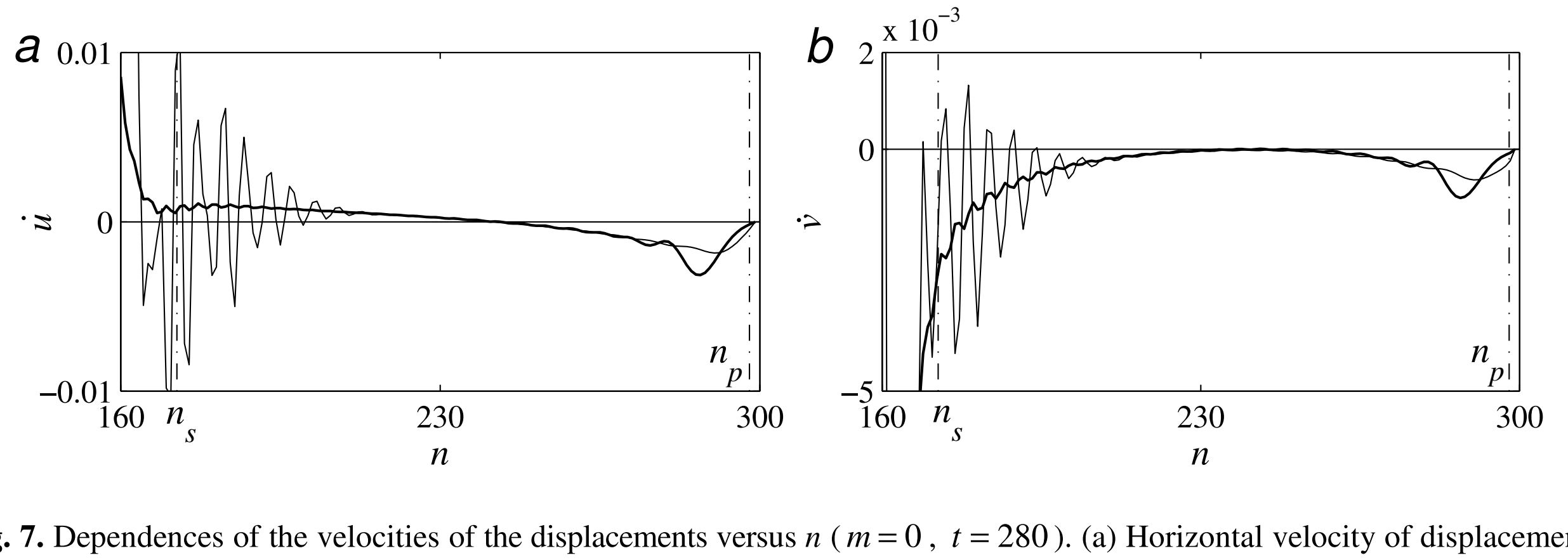

**Fig. 7.** Dependences of the velocities of the displacements versus $n$ ($m = 0$, $t = 280$). (a) Horizontal velocity of displacement $\dot{u}$, (b) vertical velocity of displacement $\dot{v}$. Thick solid lines – $\gamma = 1$. Thin solid lines – $\gamma = 1/2$.

$n_p = tc_p$, $n_s = tc_s$, $n_R = tc_R$. In Figs. 6, 8, and 10 – 13 the oscillograms of perturbations are presented for a set of the values of *n*. On those figures, the vertical lines correspond to the arrival time of the quasi-fronts of the longitudinal, shear, and Rayleigh waves (4), (7): $t_p = n/c_p$, $t_s = n/c_s$, and $t_R = n/c_R$.

In Fig. 5, we plot the displacements *u*, *v* versus the coordinate *n* on the boundary of the half-plane ($m = 0$) at the moment of time $t = 280$ ($\gamma = 1$).

As it was mentioned above, in [24] the analytical solution was given to the static Lamb problem for the plane stress state for an isotropic elastic medium. In our notation the solution, given in [24], has the following form on the boundary $m = 0$:

$$u_{n,0} = \frac{P_0}{E}\frac{(1-\sigma)}{2}, \qquad v_{n,0} = \frac{2P_0}{\pi E}\left(\ln\frac{nl}{h} + \frac{1+\sigma}{2}\right), \tag{24}$$

where *h* is a constant. If $l \to 0$, Eqs. (1) and (2) are transformed into the equations describing the plane stress state of an isotropic elastic medium with Poisson's ratio $\sigma = 1/3$. For this reason we put $\sigma = 1/3$ in (24). In Fig. 5, the dashed lines correspond to (24). In the calculations we assumed that $h = c_p t$, i.e., that *h* is the distance at which the disturbances, running at the maximal velocity in the medium, travel during time *t*.

As can be seen from Fig. 5(a), for the transient problem the horizontal displacement behind the front of the Rayleigh wave oscillates near solution (24) to the static Lamb problem. The latter is equal to $P_0(1-\sigma)/(2E)$ or, equivalently, to $P_0/(4k_1)$ which appeared as a multiple in (15). From Fig. 5(a) we see that the analytical solution (15) is consistent both qualitatively and quantitatively with the results of the numerical computations for the horizontal displacements.

For the vertical displacement *v* (see Fig. 5(b)) we observe the qualitative agreement of the numerical solution to the transient problem (1) and (2) and solution (24) to the static problem that takes place behind the front of the Rayleigh wave.

Fig. 6 shows the oscillograms of the displacements on the boundary of the block medium at the point $n = 80$ ($\gamma = 1$). From Fig. 6(a) we see that the results of the numerical calculations agree with analytical solution (15) for the horizontal displacement. From Fig. 6(b) we see that the vertical displacement is increasing in time but is not a monotonic function.

Fig. 7 shows the dependences of the velocities $\dot{u}$, $\dot{v}$ versus *n* for $t = 280$ on the boundary of the half-plane ($m = 0$) in the interval $c_s t \le n \le c_p t$. From Fig. 7(a) we see that, in the vicinity of the front of the shear wave, the amplitude of the oscillations of $\dot{u}$ for $\gamma = 1$ is significantly lower than for $\gamma = 1/2$. Similarly, from Fig. 7(b) we see that, in the vicinity of the front of the shear wave, the amplitude of the oscillations of $\dot{v}$ for $\gamma = 1$ is significantly lower than for $\gamma = 1/2$. This confirms the earlier conclusion about the strong dispersion for $\gamma < 1$.

In Fig. 8 we show the plots of the velocities $\dot{u}$, $\dot{v}$ of the displacements on the boundary of the half-plane at the point $n = 80, m = 0$ for $\gamma = 1$. In Fig. 8(b), we show the results of the same numerical calculations for $\dot{u}$ that are presented in Fig. 8(a), but in an enlarged scale along the vertical axis and the smaller time interval $t_p \le t \le t_s$. Similarly, in Fig. 8(d) we show the results for $\dot{v}$ that are presented in Fig. 8(c), but in an enlarged scale. Furthermore, in Fig. 8 dot-dashed line shows an analytical solution to the transient Lamb problem for an elastic medium that may be found in [16,23]. In the notation of our article, that solution has the form ( $x = nl/(c_p t)$, $\beta = c_p^2/c_s^2$ ):

$$R(x) = (2 - x^2\beta)^2 - 4\sqrt{(1-x^2)(1-x^2\beta)}, \quad R_1(x) = (2 - x^2\beta)^2 + 4\sqrt{(1-x^2)(1-x^2\beta)},$$

$$\dot{u}_{n,0}(t) = \begin{cases} \dfrac{2P_0 x^2\beta(x^2\beta - 2)\sqrt{(1-x^2)(x^2\beta - 1)}}{\pi\rho c_s^2 t R(x) R_1(x)}, & c_s t \le nl \le c_p t, \\ \dfrac{P_0 t\beta(x^2\beta - 2)\delta(x - x_R)}{2\rho c_s^2 x^4 R'(x)}, & nl < c_s t, \end{cases} \tag{25}$$

$$\dot{v}_{n,0}(t) = \begin{cases} -\dfrac{P_0 x^2\beta(2 - x^2\beta)^2\sqrt{1-x^2}}{\pi\rho c_s^2 t R(x) R_1(x)}, & c_s t \le nl \le c_p t, \\ -\,\text{p.v.}\left(\dfrac{P_0 x^2\beta\sqrt{1-x^2}}{\pi\rho c_s^2 t R(x)}\right), & nl < c_s t. \end{cases}$$

Here $\delta$ is the Dirac delta.

From Fig. 8 we see that the velocities of the displacements for the block medium oscillate around the solutions for the elastic medium (25). In the interval $t_p \le t \le t_s$, there is a quantitative agreement between the numerical solutions for the block medium and the analytical solutions for the elastic medium (25). Comparing the asymptotic and numerical solutions for the block medium that are shown in Figs. 8(a), 8(c), we see that asymptotic solutions (11), (12), (14), and (16) and the finite-difference solutions qualitatively agree with each other in the vicinity of the front of the Rayleigh wave and behind the front. Quantitatively, the amplitudes of the oscillations of the numerical and analytical solutions for $\dot{u}$, $\dot{v}$ differ by 15-20%, the frequencies of the oscillations differ by 18-20%. Analytical solutions (11), (14) for $\dot{u}$ coincide with each other in the vicinity of the quasi- front of the Rayleigh wave. Similarly, analytical solutions (12), (16) for $\dot{v}$ coincide with each other. For each pair of those solutions, the distinction becomes significant at a great distance from the front of the

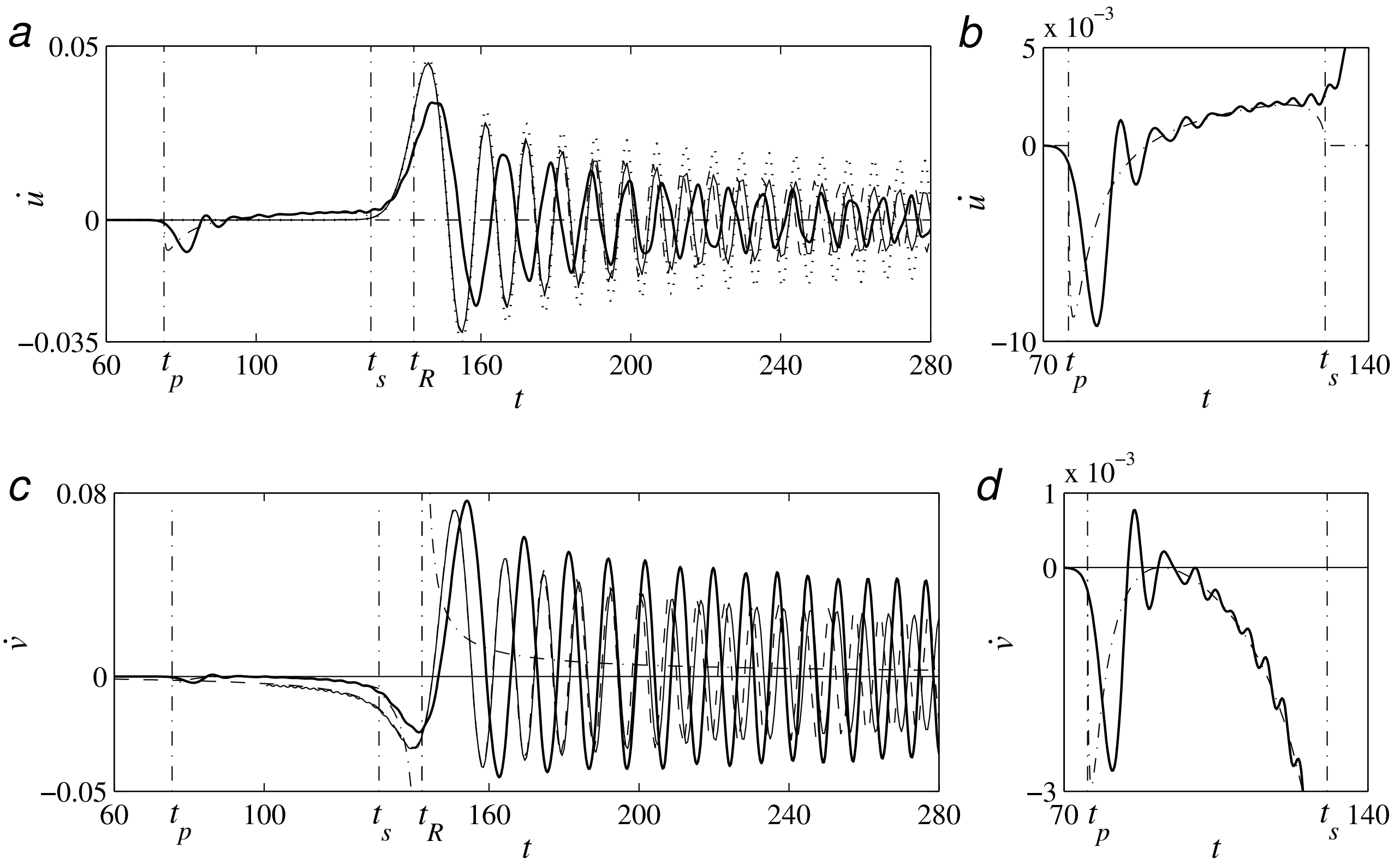

**Fig. 8.** Oscillograms of the velocities of the displacements at the point $n = 80$, $m = 0$ ($\gamma = 1$): (a), (b) horizontal velocity of the displacement $\dot{u}$, (c), (d) vertical velocity of the displacement $\dot{v}$. Thick solid lines – the results of numerical calculations. Thin solid lines – asymptotic solutions (14), (16). Dashed lines – asymptotic solutions (11), (12). Dot-dashed lines – analytical solution (25). Dotted line – asymptotic solution (18).

Rayleigh wave, and the distinction is only in frequency. The amplitudes of oscillations are the same even at large distances from the wave-front. Comparison of analytical solutions (14), (18), obtained by different methods, shows that they coincide in the vicinity of the quasi-front $nl = c_R t$. Behind the quasi-front, the amplitudes of these solutions differ essentially and solution (14) describes the results of the numerical calculations more precisely than solution (18).

In order to analyze the spectral characteristics of the oscillations of the block medium, we calculate the spectral densities of $\dot{u}$ and $\dot{v}$ determined by the following formulas:

$$G_u(\omega) = \left| \int_0^T \dot{u}_{n,m}(t) e^{-i\omega t} dt \right|, \qquad G_v(\omega) = \left| \int_0^T \dot{v}_{n,m}(t) e^{-i\omega t} dt \right|.$$

The spectral densities are calculated numerically using the Fast Fourier Transform ($T = 640$). Fig. 9 shows the plots of functions $G_u$ and $G_v$, corresponding to the oscillograms of the velocities, given in Fig. 8. In Fig. 9, the vertical lines correspond to the frequency $\omega_1 = 1$ of the short-wave perturbations in the Rayleigh wave in the block medium, where $\omega_1$ is found according to (8). From Fig. 9 we see that the spectra of the oscillations arising in the block medium under the surface load are limited by the resonance frequency of the surface waves: $\omega \le \omega_1$ .

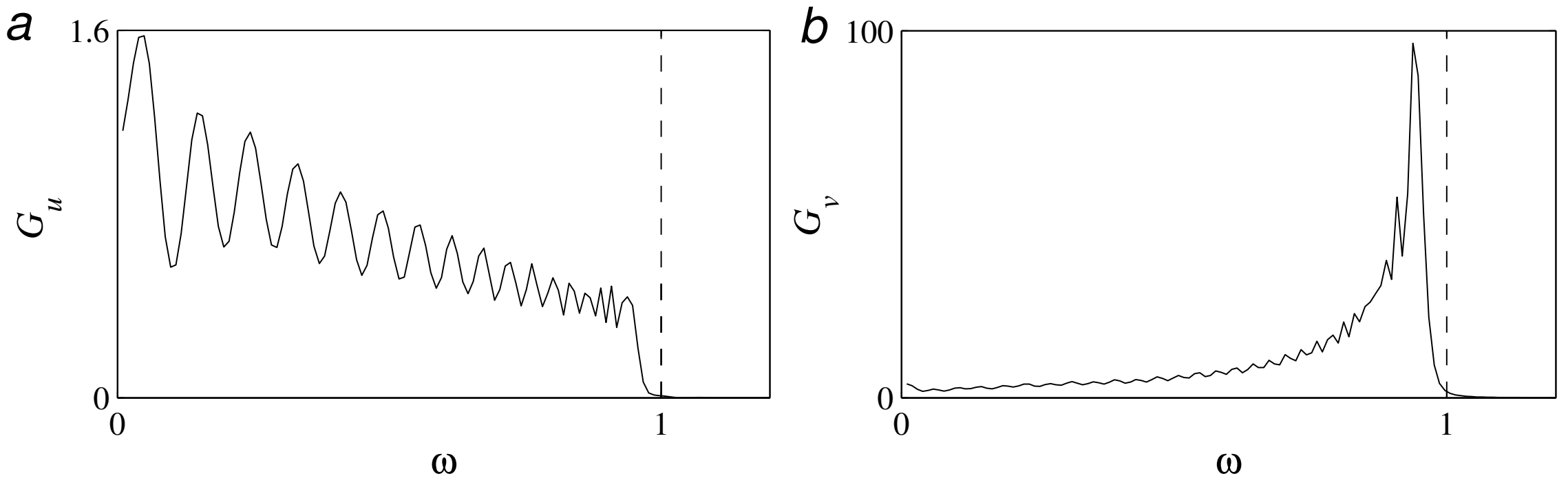

**Fig. 9.** Spectral densities of the velocities of the displacements $\dot{u}$ , $\dot{v}$ : (a) $G_u$ , (b) $G_v$ .

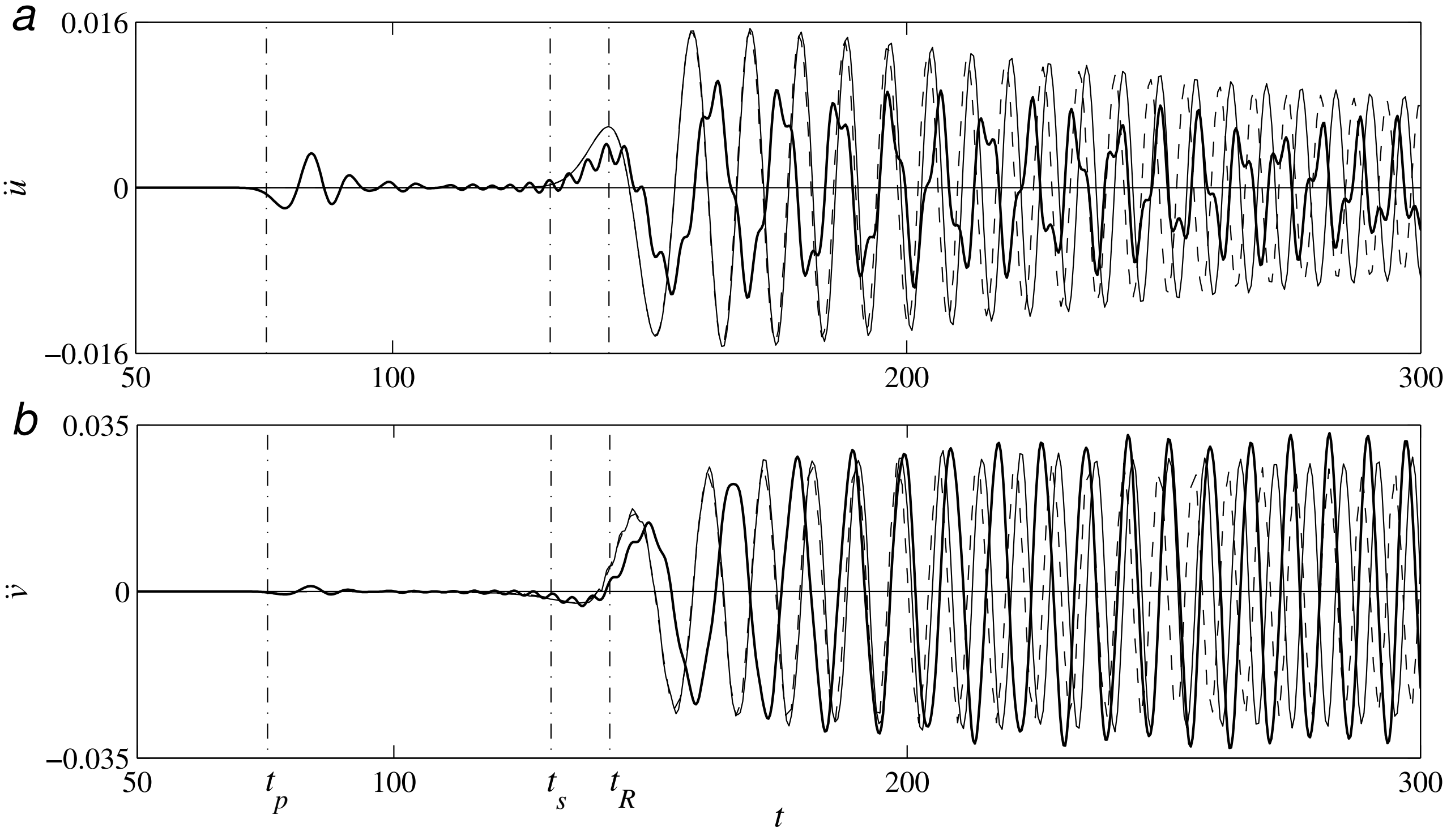

**Fig. 10.** Oscillograms of the accelerations of the displacements at the point $n = 80$ , $m = 0$ ( $\gamma = 1$ ): (a) horizontal acceleration $\ddot{u}$ , (b) vertical acceleration $\ddot{v}$ . Thick solid lines – the results of numerical calculations. Thin solid lines – asymptotic solutions (22), (23). Dashed lines – asymptotic solutions (20), (21).

Fig. 10 shows the oscillograms of the accelerations $\ddot{u}$ and $\ddot{v}$ on the boundary of the block medium at the point $n = 80$ ( $\gamma = 1$ ). From Fig. 10 we see that the frequencies of the oscillations of the numerical and asymptotic solutions for $\ddot{u}$ and $\ddot{v}$ differ by 18-20%, the amplitudes of oscillations for $\ddot{u}$ differ by 25-35%, and for $\ddot{v}$ – by 5-10%. Analytical solutions (20), (22) for $\ddot{u}$ coincide with each other in the vicinity of the quasi-front of the Rayleigh wave. Similarly, analytical solutions (21), (23) for $\ddot{v}$ coincide with each other. For each pair of these solutions, the distinction becomes significant at a great distance from the front of the Rayleigh wave, and the distinction is only in frequency. The amplitudes of oscillations are almost the same even at large distances from the wave-front.

Return to the analytical solutions obtained above. As noted above, the asymptotic solutions for the velocities and accelerations differ significantly from the results of numerical calculations, both in amplitude and frequency. Let us try to improve the asymptotic solutions. The asymptotic analysis of perturbations was carried out for $p \to 0$ and $|q| << 1$. Nevertheless, the contribution of the short-wave perturbations was also taken into account since, in formulas (11), (12), the integration was carried out over the interval $[0,\pi]$. In the neighborhood of the point $ql = \pi$, the asymptotic expression (10) for the function $\Delta\Delta_1$ used in the derivation of solutions (11), (12), (14) – (16), (20) – (23) determines the frequency of oscillations $\omega = c_R/l$ and this frequency is not equal to $\omega_1$, the value that is given by (8). In formulas (14) – (16), (22), (23) the frequency of oscillations behind the wave front depends on the coefficient $\alpha = l^2 c_R/8$ that appeared in the denominator of the expression $\kappa$. In order to make the frequency of the short-wave perturbations in the analytical solutions equal to $\omega_1$, we replace $\kappa$ by $\tilde{\kappa} = (nl - c_R t)\Big/\left(l^3 \omega_1 t/8\right)^{1/3}$. This yields:

$$\tilde{u}_{n,0}(t) \approx \frac{P_0}{4k_1}\left[\frac{1}{3} - \int_0^{\tilde{\kappa}} \mathrm{Ai}(y)dy\right]. \tag{26}$$

Differentiating (26) with respect to time, we obtain the following formula for the velocity:

$$\dot{\tilde{u}}_{n,0}(t) \approx \frac{P_0}{2k_1 l}\frac{c_R \mathrm{Ai}(\tilde{\kappa})}{(\omega_1 t)^{1/3}}. \tag{27}$$

The amplitude of oscillations in (27) differs from the amplitude of oscillations in (18) by the factor $(c_R/l\omega_1)^{1/3}$. Let us introduce this factor into (14):

$$\dot{\tilde{u}}_{n,0}(t) \approx \frac{nP_0}{2k_1 t}\frac{\mathrm{Ai}(\tilde{\kappa})}{(\omega_1 t)^{1/3}}. \tag{28}$$

Differentiating (28) with respect to time we obtain

$$\ddot{\tilde{u}}_{n,0}(t) \approx -\frac{nP_0 c_R}{k_1 l t}\frac{\mathrm{Ai}'(\tilde{\kappa})}{(\omega_1 t)^{2/3}}. \tag{29}$$

The amplitude of oscillations in (29) differs from the amplitude of oscillations in (22) by the factor $(c_R/l\omega_1)^{2/3}$.

Similarly, we correct the asymptotic solution for the vertical velocities and accelerations (more precisely, we replace $\kappa$ by $\tilde{\kappa}$, but do not change the amplitude):

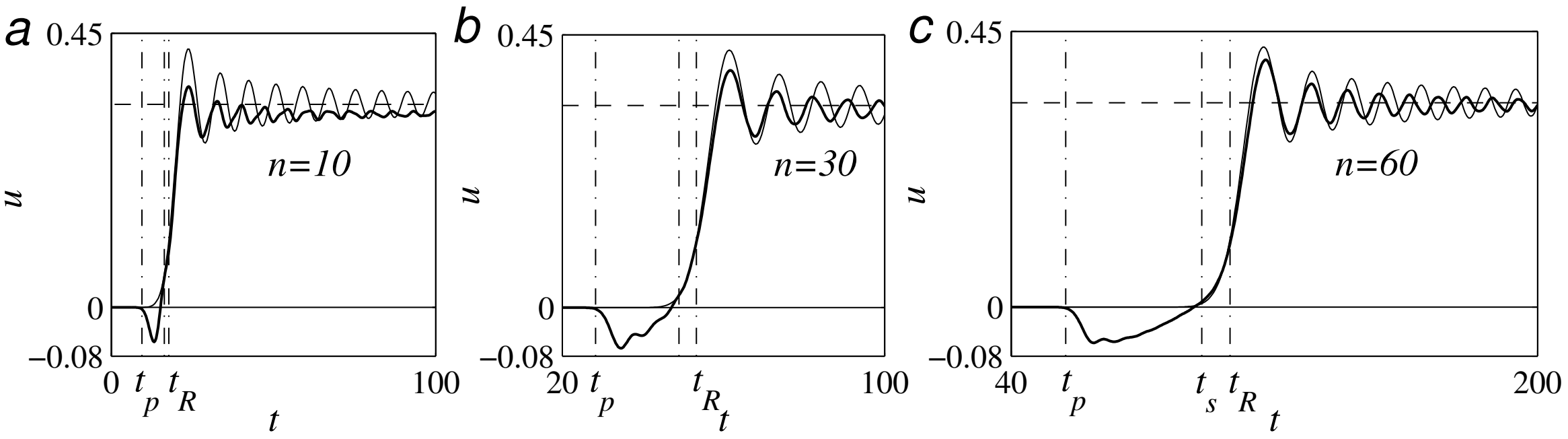

**Fig. 11.** Oscillograms of the horizontal displacement $u$ on the boundary of the block medium at the points $n = 10, 30, 60$ ( $\gamma = 1$ ). Thick solid lines – the results of numerical calculations. Thin solid lines – asymptotic solution (26). Dashed lines – the value $P_0/(4k_1)$ .

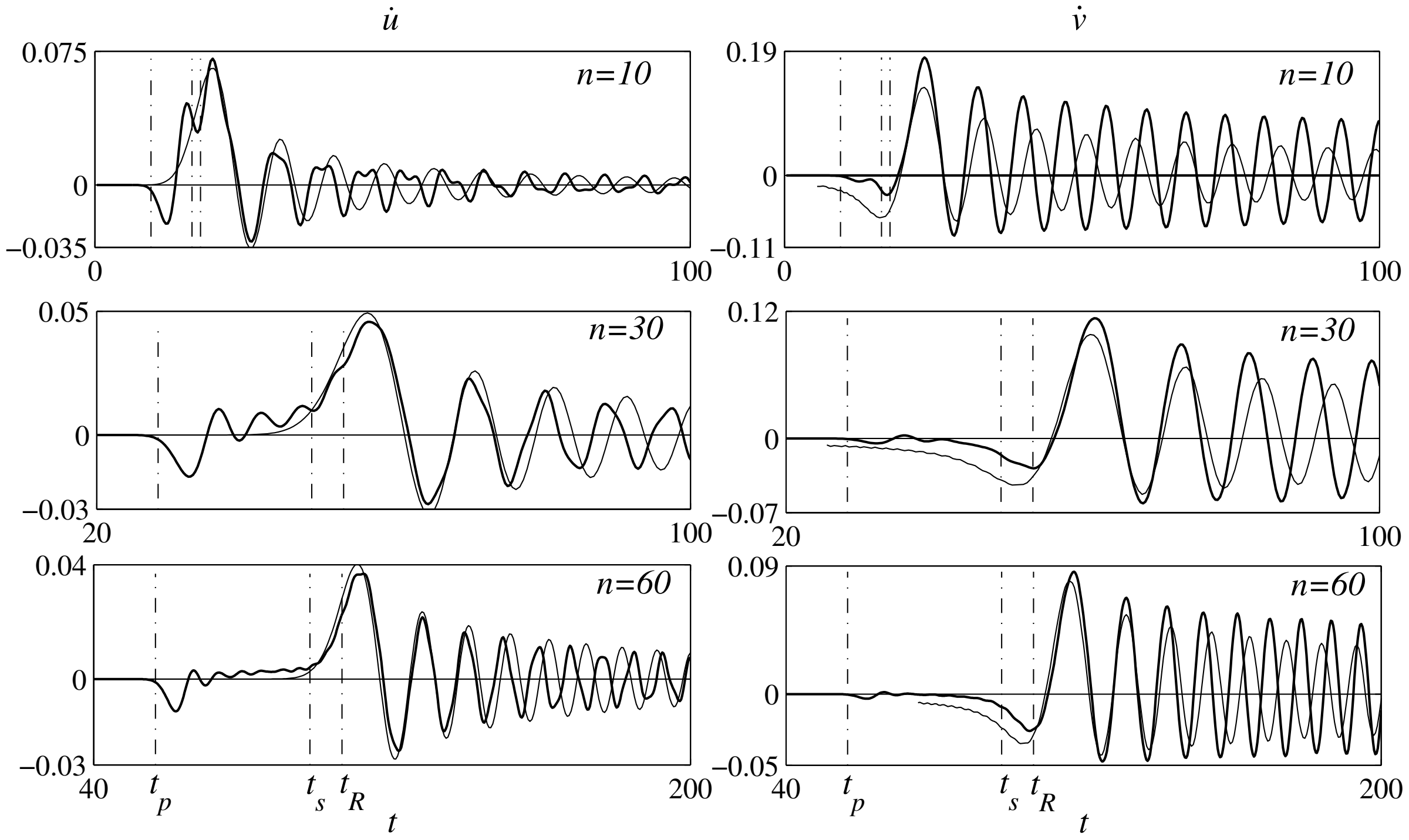

**Fig. 12.** Oscillograms of the velocities of the displacements $\dot{u}$ , $\dot{v}$ on the boundary of the block medium at the points $n = 10, 30, 60$ ( $\gamma = 1$ ). Thick solid lines – the results of numerical calculations. Thin solid lines – asymptotic solutions (28), (30). Left plots – $\dot{u}$ , right plots – $\dot{v}$ .

$$\dot{\tilde{v}}_{n,0}(t) = -\frac{P_0 l^2 \alpha_1}{4Mc_R}\frac{\mathrm{Gi}(\tilde{\kappa})}{(\alpha t)^{1/3}}, \tag{30}$$

$$\ddot{\tilde{v}}_{n,0}(t) = \frac{P_0 l^2 \alpha_1}{4M}\frac{\mathrm{Gi}'(\tilde{\kappa})}{(\alpha t)^{2/3}}. \tag{31}$$

Let us compare the new solutions (26), (28) – (31) with the results of the numerical calculations. Figs. 11–13 show the results of the numerical calculations for the horizontal displacement $u$, velocities $\dot{u}$ , $\dot{v}$ , and accelerations $\ddot{u}$ , $\ddot{v}$ , as well as analytical solutions (26), (28) – (31) at various points on the boundary of the block medium $n = 10, 30, 60$ ( $\gamma = 1$ ).

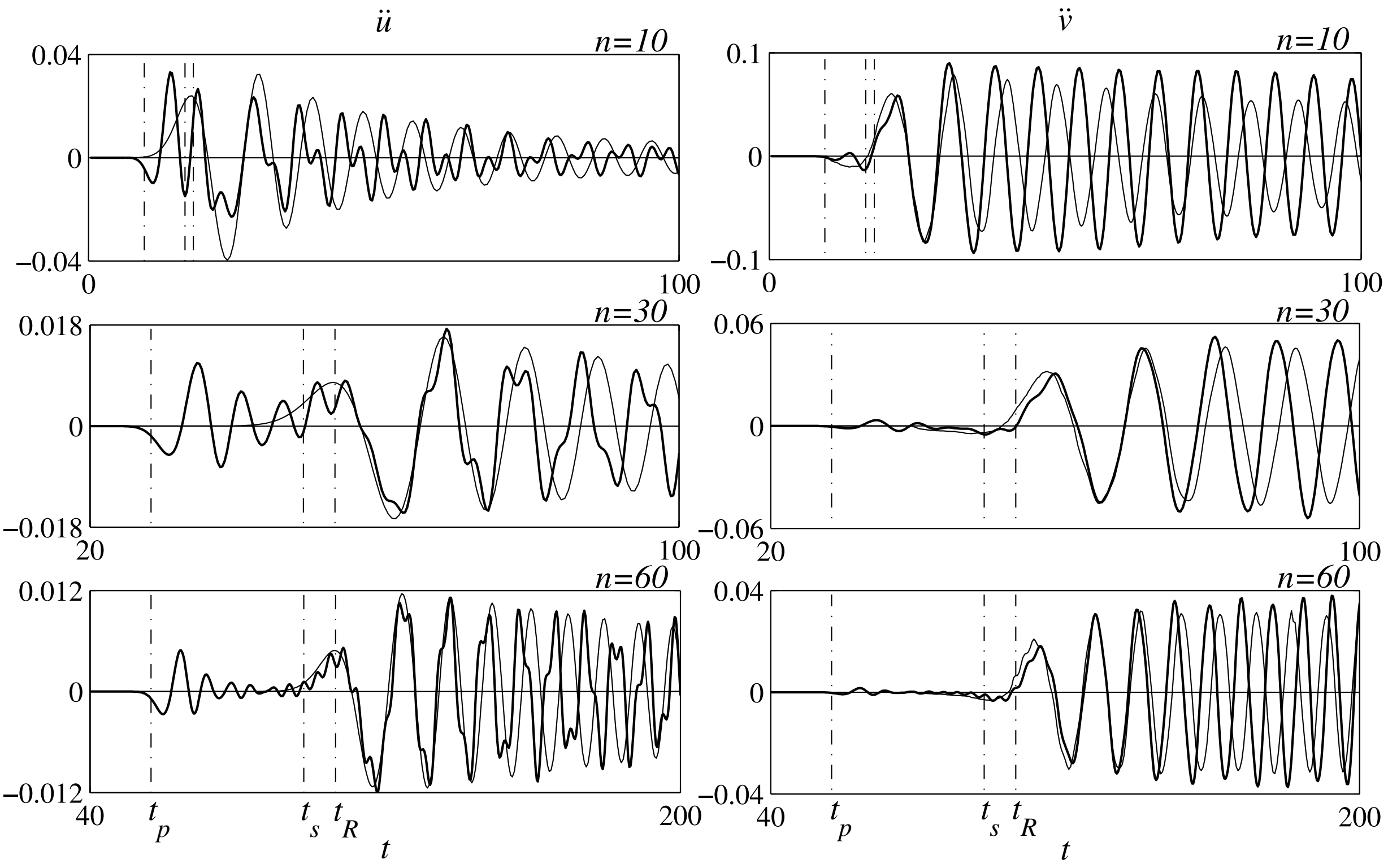

**Fig. 13.** Oscillograms of the accelerations $\ddot{u}$, $\ddot{v}$ on the boundary of the block medium at the points $n = 10, 30, 60$ ($\gamma = 1$). Thick solid lines – the results of numerical calculations. Thin solid lines – asymptotic solutions (29), (31). Left plots – $\ddot{u}$, right plots – $\ddot{v}$ .

Fig. 11 shows that, for small values of $n$ $(n < 40)$, the numerical solution for $u$ oscillates behind the front of the Rayleigh wave about a value that is smaller than the value $P_0/(4k_1)$, predicted by the analytical solution (26). And only starting from $n \approx 40$, the horizontal displacement $u$ found numerically oscillates about the value $P_0/(4k_1)$. Comparing Figs. 5, 6, 11, we also see that the frequency of the oscillations of the numerical solution for $u$ corresponds to the frequency of the analytical solution (26) better than to the frequency of the analytical solution (15).

Fig. 12 shows that the conformity of the frequencies and maximal amplitudes of the oscillations of $\dot{u}$, $\dot{v}$ near the quasi-front of the Rayleigh wave for the finite-difference and analytical (28), (30) solutions becomes better as $n$ increases. Note that, despite the fact that the analytical solutions (28), (30) were obtained under the assumption that $t \to \infty$ (or, equivalently, that $n \to \infty$), they describe the numerical solution at a finite distance from the place of impact with a reasonable accuracy, namely, for the horizontal velocity of displacement $\dot{u}$ – starting from $n = 10$, and for the vertical velocity of displacement $\dot{v}$ – starting from $n = 30$.

In Fig. 13 we see that the conformity of the amplitudes and frequencies of the oscillations of the horizontal acceleration $\ddot{u}$ near the quasi-front of the Rayleigh wave for analytical (29) and numerical solutions is achieved starting from $n = 30$. For the vertical acceleration $\ddot{v}$, the analytical (31) and numerical solutions match for $n$ starting from $n = 10$.

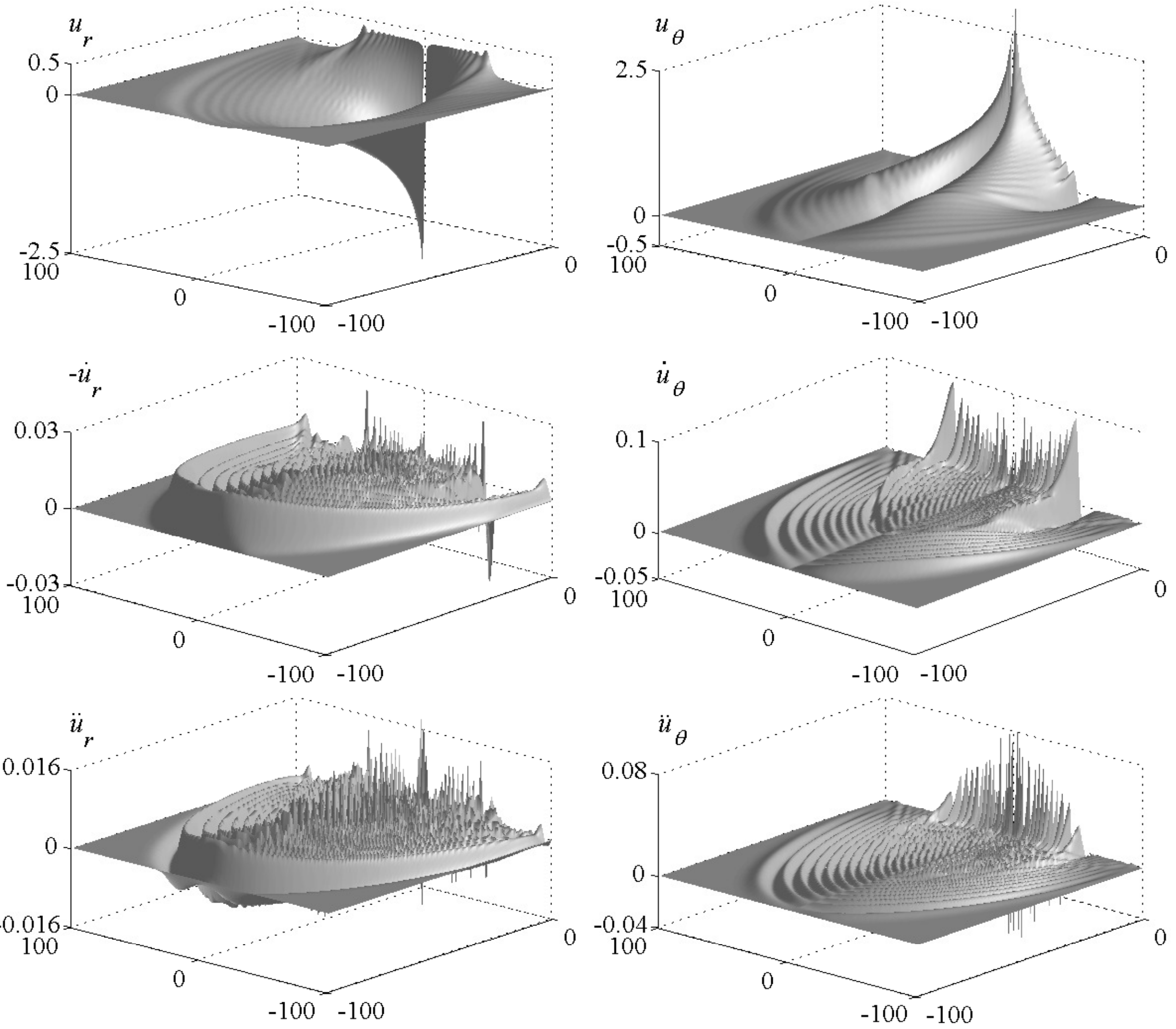

**Fig. 14.** Plots of tangential $u_\theta$ and radial $u_r$ displacements and their velocities $\dot{u}_\theta$, $\dot{u}_r$ and accelerations $\ddot{u}_\theta$, $\ddot{u}_r$ on the half-plane at the moment of time $t = 94$ ( $\gamma = 1$ ).

Thus, we have shown that the corrected formulas (26), (28) – (31) describe the results of the numerical calculations with higher accuracy than formulas (11), (12), (14) – (16), (20) – (23).

In Fig. 14, we plot tangential $u_\theta$ and radial $u_r$ displacements

$$u_\theta = sign(n)\frac{v_{n,m}n + u_{n,m}m}{\sqrt{n^2 + m^2}}, \qquad u_r = \frac{v_{n,m}m + u_{n,m}n}{\sqrt{n^2 + m^2}},$$

and their velocities $\dot{u}_\theta$, $\dot{u}_r$ and accelerations $\ddot{u}_\theta$, $\ddot{u}_r$ on the half-plane at the moment of time $t = 94$. The fronts of the longitudinal, shear and Rayleigh waves are clearly visible. The traveling longitudinal wave interacts with the free surface and generates a head shear wave with a straight front. The front of this wave is visible on the plot of function $\dot{u}_\theta$.

## 5. Conclusion

The study of the dynamic behavior of the two-dimensional model of the block medium under a vertical impact load on the half-plane (the Lamb problem) reveals that the presence of the structure in the medium causes a change in its behavior in comparison with that predicted by the continuous model obtained by the averaging of the mechanical properties of the block medium. The difference between the block medium and the continuous medium is particularly large in the vicinity of the front of the Rayleigh wave, namely, there appear high-frequency oscillations of disturbances which are absent in the continuous medium. In the area on the boundary of the block medium, contained between the fronts of the longitudinal and shear waves, there is a qualitative and quantitative agreement between the results for the block and homogeneous media. Dispersion characteristics of waves propagating in the half-space filled by the block medium are more complicated in comparison with the elastic medium. Longitudinal, shear and Rayleigh waves in the block medium propagate with dispersion. There is no such effect in the elastic medium. The spectrum of the high-frequency oscillations is bounded by the resonance frequency of the surface waves of the block medium and is determined by its structural properties.

Asymptotic solutions to the transient problem are obtained that describe the behavior of the perturbations on the boundary of the block medium in the vicinity of the Rayleigh wave at large time since the beginning of the process or at the large distance from the loading point.

As a result of the study of the asymptotic solutions it is shown that:

**(a)** behind the front of the Rayleigh wave, the horizontal displacement oscillates around a constant value which coincides with the static solution to the Lamb problem;

**(b)** the amplitudes of the velocities of the displacements $\dot{u}, \dot{v}$ in the vicinity of the quasi-front of the Rayleigh wave decrease with time (or distance) as $t^{-1/3}$ (or $n^{-1/3}$);

**(c)** the amplitudes of the accelerations $\ddot{u}, \ddot{v}$ in the vicinity of the quasi-front of the Rayleigh wave decrease with time (or distance) as $t^{-2/3}$ (or $n^{-2/3}$);

**(d)** for functions $u, \dot{u}, \dot{v}, \ddot{u}, \ddot{v}$, the width of the quasi-front expands with time (or distance) as $t^{1/3}$ (or $n^{1/3}$).

The comparison of the finite-difference and analytical solutions reveals that the long-wave perturbations make the main contribution to the wave process in the vicinity of the quasi-front of the Rayleigh wave. Behind the front the main contribution is made by the short-wave oscillations. The two solutions match with each other at a finite interaction time, or what is the same, at a finite distance from the loading point.

Thus, we have shown that, in order to describe the dynamic behavior of the block medium, we can simulate its deformation as a motion of rigid blocks separated by compressible intermediate layers (the pendulum approximation).

In the process of deriving the asymptotic solutions we obtain a new approximate representation of the Lommel function $s_{0,k}$ for $k >> 1$ in terms of the Scorer function $\mathrm{Gi}$.

## Acknowledgements

This research was supported in part by the Russian Foundation for Basic Research (grant no. 11-05-00369) and by the Ministry of Education and Science of the Russian Federation.